\documentclass[aps,prd,twocolumn,floatfix,showpacs,superscriptaddress,nofootinbib]{revtex4-2}
\usepackage{amsmath,amsfonts,amssymb,bm}
\usepackage{graphicx}
\usepackage{color}
\usepackage{subfigure}
\usepackage{multirow}
\usepackage{textcomp}
\usepackage{slashed}
\usepackage[bottom]{footmisc}
\usepackage[T1]{fontenc}
\definecolor{purple}{rgb}{0.5,0,0.5}
\definecolor{blue}{rgb}{0.0,0,1.0}
\usepackage[colorlinks=true, pdfstartview=FitV, linkcolor=purple, citecolor= purple, urlcolor=blue]{hyperref}

\newcommand{\beq}{\begin{equation}}
\newcommand{\eeq}{\end{equation}}
\newcommand{\ba}{\begin{array}}
\newcommand{\ea}{\end{array}}
\newcommand{\bea}{\begin{align}}
\newcommand{\eea}{\end{align}}
\newcommand{\bi}{\begin{itemize}}
\newcommand{\ei}{\end{itemize}}
\newcommand{\ben}{\begin{enumerate}}
\newcommand{\een}{\end{enumerate}}
\newcommand{\bc}{\begin{center}}
\newcommand{\ec}{\end{center}}
\newcommand{\bl}{\begin{flushleft}}
\newcommand{\el}{\end{flushleft}}
\newcommand{\br}{\begin{flushright}}
\newcommand{\er}{\end{flushright}}

\begin{document}

\preprint{APS/123-QED}

\title{Contact interaction treatment of $\mathcal{V}\to\mathcal{P}\gamma$ for light-quark mesons}

\author{Yehan Xu}
\email{xuyehan@mail.nankai.edu.cn}
\affiliation{%
 School of Physics, Nankai University, Tianjin 300071, China 
}%
\author{M. Atif Sultan}%
\email{atifsultan.chep@pu.edu.pk}
\affiliation{%
 School of Physics, Nankai University, Tianjin 300071, China 
}%
\affiliation{  Centre  For  High  Energy  Physics,  University  of  the  Punjab,  Lahore  (54590),  Pakistan}
\author{Kh\'epani Raya}
\email{khepani.raya@dci.uhu.es}
\affiliation{Department of Integrated Sciences and Center for Advanced Studies in Physics, Mathematics and Computation, University of Huelva, E-21071 Huelva, Spain.}
\author{Lei Chang}
\email{leichang@nankai.edu.cn}
\affiliation{%
 School of Physics, Nankai University, Tianjin 300071, China 
}%

\date{\today}
\begin{abstract}

The $\mathcal{V}\to\mathcal{P}\gamma$ and $\eta(\eta^\prime) \to \gamma\gamma$ decays are evaluated within a Dyson-Schwinger and Bethe-Salpeter equations framework (here $\mathcal{V}=\{\rho^{\pm},K^{\star\pm},\phi\}$ and $\mathcal{P}=\{\pi^{\pm},K^{\pm},\eta,\eta^{\prime}\}$).  The so-called impulse approximation (IA) is employed in the computation of the decay constants involved and decay widths, and so in the estimation of the associated charge and interaction radii. For their part, the required propagators and vertices stem from a contact interaction model, embedded within a beyond rainbow-ladder (RL) truncation that accounts for the typical ladder exchanges, quark anomalous magnetic moment, as well as the non-Abelian anomaly. While the examined transitions produce decay widths plainly compatible with the available experimental data, those processes involving the $\eta-\eta'$ mesons highlight the incompleteness of the IA when considering beyond RL effects in the interaction kernels.
\end{abstract}

\maketitle
\section{\label{introduction}introduction}
The Standard Model (SM) of particle physics is an immensely successful theory that unifies the description of electromagnetic, weak and strong forces in terms of elementary particles. Among its components, quantum chromodynamics (QCD) is intended to describe the properties of hadronic matter. Nonetheless, enormous challenges arise due to the peculiarities of the strong interactions\,\cite{Marciano:1977su,Marciano:1979wa}. First of all, QCD's fundamental degrees of freedom (quarks and gluons) cannot be identified in isolation; what we get to detect are composite neutral-colored objects (hadrons) with typical sizes of the order of $1$\,fm. This phenomenon is commonly dubbed as confinement. Secondly, whereas asymptotic freedom causes the coupling characterizing the strong interactions become small in the large-energy regime, hence allowing a perturbative treatment of QCD, daily life phenomena occur in the opposing end\,\cite{Roberts:2021nhw,Ding:2022ows}. Mass generation is the prime example of such: through their own mechanisms, dynamical mass generation is present in both matter and gauge sectors\,\cite{Deur:2023dzc,Papavassiliou:2022wrb,Cui:2019dwv,Binosi:2014aea}. This orchestrates the so called emergence of hadron masses (EHM), which produces almost all of the mass of the visible universe. Understanding the non-perturbative facets of QCD hence implies elucidating the hadron spectrum and structure,  tracing down its connection with EHM and confinement.

Light pseudoscalar mesons (PS), namely $\mathcal{P}=\{\pi^{\pm},K^{\pm},\eta,\eta^{\prime}\}$, play a special role in this context. Their origins are tightly connected with the breaking of chiral symmetry, and so EHM\,\cite{Horn:2016rip,Raya:2024ejx}. With the exception of the $\eta'$, these are the lightest bound-states produced by QCD, \emph{i.e.} the pseudo Nambu-Goldstone (NG) bosons of dynamical chiral symmetry breaking (DCSB); consequently, becoming massless and structurally identical in the absence of the Higgs mechanism (HBM). The influence of the so-called Abelian and non-Abelian anomalies sets the $\eta-\eta'$ system apart, allowing the latter to remain massive in the absence of HBM, but intertwining its properties with those of the $\eta$,\,\cite{Fritzsch:1976qc,Gilman:1987ax}. These facts strongly suggest that the structural differences amongst hadrons are profoundly influenced by the interplay between the different mass generation mechanisms and anomalies of the SM. Complementary insights are obtained from light vector mesons (VM), $\mathcal{V}=\{\rho^{\pm},K^{\star\pm},\phi\}$, \emph{e.g.} Refs.\,\cite{Xu:2019ilh,Kim:2022wkc,Xu:2024vkn,Kaur:2024iwn}. In a rather simplistic picture, these systems can be regarded as the lowest-lying spin excitations of their PS partners; and, contrary to the latter, VM exhibit a mass budget that closely resembles that of the nucleon\,\cite{Ding:2022ows}. Contrasting the properties of VM with those of the PS and nucleons would expose once again the effects of the mass generation mechanisms, while also unveiling the role of the spin\,\cite{Goeke:2001tz,Lorce:2022jyi,Sheng:2022ffb}. Furthermore, being characterized by the same quantum numbers as the photon, VM provide a direct link between the strong interaction and the electromagnetic force, becoming instrumental in the study of space- and time-like electromagnetic form factors (FFs) and numerous decay processes\,\cite{Bauer:1977iq,GarciaGudino:2015ocw,Xu:2021mju,Burkert:2022hjz,Cao:2023rhu}. 

The present analysis focuses on the description of the $\mathcal{V}\to\mathcal{P}\gamma$ spin-flip transitions, as well as the related $\eta(\eta^\prime) \to \gamma\gamma$ decays. These processes not only allow us to pin down structural changes generated by the electromagnetic interaction, but are also linked to the aforementioned SM anomalies. For instance, the so-called Abelian anomaly (AA) contributes significantly to the the $\rho \to \gamma \pi$ and $\{\pi,\,\eta,\,\eta'\}\to\gamma \gamma$ decay amplitudes, whereas the flavor mixing effects produced by the non-Abelian anomaly (NAA) introduces additional subtleties in the $\phi\to\eta(\eta^{\prime})\gamma$ case. Experimentally, the $\mathcal{V}\to\mathcal{P}\gamma$ decay channels play an important role in vector electromagnetic processes, and there exists a number of measurements of the radiative decay widths and transition FFs for both space- and time-like energy regions (see \textit{e.g.} Refs.~\cite{Dzhelyadin:1980tj,NA60:2009una,NA60:2016nad,Adlarson:2016hpp,Achasov:2000ne,KLOE-2:2014hfk}). From a theoretical perspective, available investigations include the early predictions based upon $SU(3)$ flavor symmetry~\cite{Glashow:1963zz}, quark models~\cite{Anisovich:1965fkk,Becchi:1965zza}, chiral effective Lagrangians~\cite{Gomm:1984at,Klingl:1996by,Benayoun:1999fv,Chen:2013nna,Kimura:2016xnx}, phenomenological Lagrangians~\cite{Durso:1987eg},  QCD sum rules~\cite{Zhu:1998bm}, and lattice QCD~\cite{Woloshyn:1986pk,Crisafulli:1991pn,Owen:2015fra,Alexandrou:2018jbt}. With respect to radiative decays involving $\eta-\eta'$ mesons, numerous studies have been carried out over a large time span; a rather short but representative sample is found in Refs.\,\cite{Fritzsch:1976qc,Gilman:1987ax,Feldmann:1997vc,Feldmann:1998vh,Feldmann:1998sh,Escribano:2005qq,Escribano:2015nra}.

Investigating these processes can be effectively achieved using the Dyson-Schwinger equations (DSEs) and Bethe-Salpeter equations (BSEs) framework. This approach captures the non-perturbative traits of QCD and incorporates all the relevant symmetries in a natural manner. Consequently, it is capable of providing a sensible description of the static and structural properties of hadrons, maintaining a traceable connection with QCD (see \emph{e.g.} Refs.\,\cite{Qin:2020rad,Eichmann:2016yit}). Amongst many explorations, the $\gamma^* \rho \to \pi$ and $\gamma^{(*)}\gamma^{(*)}\to \{\pi^0,\eta,\eta',\eta_c,\eta_b\}$ transition FFs have been accurately described using sophisticated interaction kernels,\,\cite{Maris:2002mz}, and through simpler approaches that rely on the implementation of a symmetry-preserving vector$\times$vector contact interaction model (CI),\,\cite{Gutierrez-Guerrero:2010waf,Roberts:2010rn}. The latter provides an accessible computational framework that preserves crucial features of QCD, such as confinement and DCSB. Moreover, it offers valuable points of comparisons, as the origins of the model's successes and deficiencies are plainly identifiable.  Hence, the CI has been used to address an endless number of hadronic properties: mass spectrum, form factors, parton distributions, among others. The growing catalog of studies within the CI encompasses, but it is not limited to, Refs.~\cite{Gutierrez-Guerrero:2010waf,Roberts:2010rn,Roberts:2011wy,Roberts:2011cf,Wilson:2011aa,Chen:2012qr,Segovia:2014aza,Serna:2017nlr,Raya:2023wye,Raya:2021pyr,Gutierrez-Guerrero:2021rsx,Yin:2019bxe,Cheng:2022jxe,Xing:2022mvk,Xing:2023eed,Dang:2023ysl,Xing:2024bpj,Ahmad:2020jzn,Wang:2013wk,Sultan:2024hep}.

In view of these facts, we shall employ a CI model to investigate the $\mathcal{V}\to\mathcal{P}\gamma$ and $\eta(\eta^\prime) \to \gamma\gamma$ radiative decays. Our analysis extends the convenional CI framework by considering a beyond rainbow-ladder (RL) truncation scheme. Firstly, we employ the interaction kernel put forward in Ref.\,\cite{Xing:2021dwe}, which incorporates the effects of the quark anomalous magnetic moment (AMM). This is known to favor the description of the vector channels and turns to be vital  for a satisfactory explanation of the $\gamma^* \gamma \to \pi^0$ and $\gamma \to 3\pi$ anomalous processes\,\cite{Dang:2023ysl,Xing:2024bpj}. The two-body kernel is further augmented to account for the NAA, in order to produce the anticipated flavor mixing and mass splittings required for the $\eta-\eta'$ system. The structure of this kernel was derived in Ref.\,\cite{Bhagwat:2007ha}, and has been succesfully applied in extensive numerical calculations of two-photon transition FFs,\,\cite{Ding:2018xwy,Raya:2019dnh}, and within the CI model\,\cite{Zamora:2023fgl}. Finally, it is worth pointing out that our treatment of the CI follows the regularization procedure introduced in Ref.~\cite{Xing:2022jtt}. Such a scheme handles the quadratic and logarithmically divergent integrals in a manner that clearly demonstrates the preservation of essential symmetries, such as the Ward-Green-Takahashi identities (WGTIs).

The manuscript is organized as follows: In Sec.\,\ref{sec:formalism}, we introduce the general aspects of the CI model, including the DSE for the quark propagator, and the corresponding meson and quark-photon vertex BSEs. With the exception of $\eta-\eta'$ mesons, some static properties of light mesons under scrutiny are also shown in this section. Our treatment of the latter is addressed in Sec.\,\ref{sec:eta}. This describes the structure of the NAA kernel and Bethe-Salpeter amplitudes (BSAs) for the mixed system, as well as the computed masses, decay constants, and mixing angles. In Sec.\,\ref{sec:raddecay}, we discuss the computation of the $\mathcal{V}\to\mathcal{P}\gamma$ and $\eta(\eta^\prime) \to \gamma\gamma$ radiative decays within the impulse approximation (IA). The presentation and discussion of results also takes place in this section. Finally, a brief summary is provided in Sec.\ref{sec:summary}.

\section{The contact interaction model}\label{sec:formalism}

\subsection{Quark propagator}\label{sec:A}
The $f$-flavored fully-dressed quark propagator might be expressed in a general manner as
\begin{equation}
    \label{eq:quarkDressed}
    S_f(p)=Z_f(p^2)[i\gamma \cdot p + M_f(p^2)]^{-1}\,,
\end{equation}
in such a way that the scalar functions $Z_f(p^2)$ and $M_f(p^2)$ (this typically denoted as mass function) are obtained from the corresponding DSE for the quark propagator, namely:
\begin{equation}
\label{eq:quarkPropGen}
    S^{-1}_f(p)=[i \gamma \cdot p+m_f]^{-1}+\int_{q} \mathcal{K}_f^{(1)}(q,p)
    S_f(q)\;;
\end{equation}
here $m_f$ denotes the bare current quark mass; $\int_q:=\int\frac{d^4q}{(2\pi)^4}$ stands for a Poincar\'{e} covariant intergral; and $\mathcal{K}^{(1)}$ represents the 1-body interaction kernel, which reads:
\begin{equation}
\label{eq:K1}
    \mathcal{K}^{(1)}(q,p)=\frac{4}{3} g^2 D_{\mu\nu}(p-q)\gamma_\mu \otimes \Gamma_\nu^f(q,p)\;,
\end{equation}
where $g$ is the Lagrangian coupling constant; $D_{\mu\nu}$ and $\Gamma_\nu$ are the fully-dressed gluon propagator and quark-gluon vertex (QGV), respectively, which obey their own DSEs.  For the sake of simplicity, we omitted the renormalization constants. Since $\mathcal{K}_f^{(1)}(q,p)$ relates the quark propagator with higher order Green functions \emph{ad infinitum}, an infinite system of coupled integral equations is formed. A tractable problem arises once a truncation scheme is specified. Thus, let us consider the interaction kernel introduced in\,\cite{Gutierrez-Guerrero:2010waf,Roberts:2010rn}, that is:
\begin{equation}
\label{eq:K1CI}
    \mathcal{K}_f^{(1)}(q,p)\to \mathcal{K}_f^{R}:= \frac{4}{3m_G^2}\delta_{\mu\nu}\gamma_\mu \otimes \gamma_\nu\;,
\end{equation}
in which $m_G$ is an infrared mass scale. This corresponds to the so-called \emph{rainbow} approximation, embedded within the CI model. Such a scheme features a momentum-independent 1-body interaction kernel, which translates into a rather simple quark DSE:
\begin{equation}
\label{eq:quarkPropCI}
    S_f^{-1}(p)=[i \gamma \cdot p+m_f]^{-1}+\frac{4}{3 m_G^2} \int_{q}  \gamma_\mu S_f(q) \gamma_\mu\;.
\end{equation}
Among other implications, an adequate treatment of the arising divergences yields a convenient expression for the quark propagator,
\begin{equation}
    \label{eq:quarkPropCIgen}
    S^{-1}_f(p)=i\gamma\cdot p + M_f\,,
\end{equation}
where the momentum-independent mass function $M_f$ plays the role of a constituent quark mass, enhanced by the effects of DCSB, and obtained from: 
\begin{equation}
M_f=m_f+\frac{M_f}{3\pi^2 m_{G}^{2}}\int_{0}^{\infty}ds\frac{s}{s+M_f^2}\;.
\end{equation}
The above integral is divergent and its form is typical of the CI model. Thus, it turns out convenient to define
\begin{equation}\begin{aligned}
I_{-2\alpha}(\mathcal{M}^{2})& :=\int_{q}\frac{1}{(q^{2}+\mathcal{M}^{2})^{\alpha+2}}\,,  \\
I_{-2\alpha}^{\mu\nu}(\mathcal{M}^2)&:=\int_q\frac{q_{\mu}q_{\nu}}{(q^2+\mathcal{M}^2)^{\alpha+3}}\,.
\end{aligned}\end{equation}
Employing the symmetry-preserving regularization procedure introduced in Ref.~\cite{Xing:2022jtt}, these integrals are regularized as follows: 
\begin{equation}\begin{aligned}
I_{-2\alpha}(\mathcal{M}^{2})\to I_{-2\alpha R}(\mathcal{M}^{2})& =\int_{\tau_{uv}^{2}}^{\tau_{ir}^{2}}d\tau\frac{\tau^{\alpha-1}}{\Gamma(\alpha+2)}\frac{e^{-\tau\mathcal{M}^{2}}}{16\pi^{2}},\\
I_{-2\alpha}^{\mu\nu}(\mathcal{M}^2)\to I_{-2\alpha R}^{\mu\nu}(\mathcal{M}^2)&=\frac{\Gamma(\alpha+2)}{2\Gamma(\alpha+3)}\delta_{\mu\nu}I_{-2\alpha R}(\mathcal{M}^2)\,,
\end{aligned}\end{equation}
where $\tau_{ir}=1/\Lambda_{ir}$ and $\tau_{uv}=1/\Lambda_{uv}$ act as infrared (IR) and ultraviolet (UV) cutoffs, respectively. A nonzero value of $\Lambda_{ir}\sim \Lambda_{\text{QCD}}$ produces a picture compatible with confinement by ensuring the absence of quark production thresholds, whereas the presence of $\Lambda_{uv}$ is mandatory within the regularization scheme and plays a dynamical role. 

Thus, from now on, we adopt the parameters used in Ref.~\cite{Xing:2022jtt}, which are typical for a description of systems containing $l=u/d$ and $s$ quarks\,\cite{Gutierrez-Guerrero:2021rsx,Yin:2019bxe}. These are captured in Table\,\ref{tab:paramsGap}, along with the produced constituent quark masses. The approach to the two-body BSE is presented in the following section.

\begin{table}[ht!]
\caption{\label{tab:paramsGap} CI model parameters entering the quark DSE, and computed constituent quark masses. Herein we employ the isospin symmetric limit $m_{u/d}=m_l$. Mass units in GeV.}
\begin{tabular}{c c c c c|c c}
\hline
    $m_G$ & $\Lambda_{ir}$ & $\Lambda_{uv}$ & $m_l$ & $m_s$ & $M_l$ & $M_s$ \\
    \hline 
    $0.132$ & $0.24$ & $0.905$ & $0.007$ & $0.17$ & $0.368$ & $0.533$\\
\hline
\end{tabular}
\end{table}

\subsection{Meson Bethe-Salpeter equation}
\label{sec:BSE1}

The BSA characterizing the internal structure of a meson is obtained via the corresponding BSE, which reads:
\begin{equation}
    \label{eq:BSEGen}
    \Gamma_H(p;P) = \int_q \mathcal{K}^{(2)}(q,p;P) \chi_H(q;P)\,.
\end{equation}
Herein, $H$ labels the type of meson; $\Gamma_H$ is the meson BSA, such that $\chi_H$ corresponds to the Bethe-Salpeter wavefunction (BSWF),
\begin{equation}
\label{eq:BSWFdef}
    \chi_H(q;P) = S_{\bar{g}}(q)\Gamma_H(q;P)S_f(q-P)\, ,
\end{equation}
where $f/\bar{g}$ the quark/antiquark flavors, respectively. The kinematic variables $p,\,q$ are the relative momentum between the valence quark-antiquark and $P$ is the total momentum of the meson, implying $P^2=-m_H^2$ ($m_H$ is the meson mass). Naturally, $\mathcal{K}^{(2)}(q,p;P)$ is the quark-antiquark scattering kernel, which expresses all the interactions that could take place within the meson. Self-consistency demands the 1-body and 2-body kernels, $\mathcal{K}_f^{(1)}$ and $\mathcal{K}^{(2)}$, to be interwinded with each other\,\cite{Chang:2009zb,Qin:2020jig}.  The simplest symmetry-preserving construction is:
\begin{equation}
    \label{eq:K2CIRL}
    \mathcal{K}^{(2)}(q,p;P)\to \mathcal{K}^{L}:=-\frac{4}{3m_G^2}\delta_{\mu\nu}\gamma_\mu \otimes \gamma_\nu=-\mathcal{K}_f^{R}\,,
\end{equation}
which corresponds to the so-called \emph{rainbow-ladder} (RL) truncation in the CI model. Here we go one step further and consider the extension proposed in Ref.\,\cite{Xing:2021dwe}. In this case, the 2-body kernel reads:
\begin{eqnarray}
    \mathcal{K}^{(2)}(q,p;P)&\to& \mathcal{K}^L + \mathcal{K}^A\,,\\
    \mathcal{K}^A&:=&\frac{4}{3m_G^2} \xi_{A} \tilde{\Gamma}_{j}\otimes \tilde{\Gamma}_{j}\;,
\end{eqnarray}
with $\tilde{\Gamma}_{j} = \{ \mathbb{I},\;\gamma_5,\;i/\sqrt{6}\sigma_{\mu\nu}\}$, and $\xi_{A}$ being a strength parameter that controls the relative weight between the RL and non-ladder (NL) contributions. So, the corresponding BSE reads:
\begin{eqnarray}
\label{eq:CIinMRL}
\Gamma_H(P)=-\frac{4}{3m_{G}^{2}}\int_{q}\left[\gamma_{\mu}\chi_H(P)\gamma_{\mu} - \xi_{A} \tilde{\Gamma}_{j}\chi_H(P)\tilde{\Gamma}_{j}\right]\,.\hspace{0.6cm} 
\end{eqnarray}
Within the present context, the meson BSAs  are independent of relative momentum of the quarks, and exhibit a compact structure:
\begin{eqnarray}
\label{eq:BSApi}
\Gamma_{PS}(P) &=& \gamma_5 \left[ i E_{PS}(P) + \frac{\gamma \cdot P}{\bar{M}} F_{PS}(P) \right]\;,\\
\label{eq:BSArho}
\Gamma^{V}_{\mu}(P)&=&\gamma_{\mu}^{T}E_{V}(P)+\frac{1}{\bar{M}}\gamma_\mu^A F_{V}(P)\;,
\end{eqnarray}
where $\gamma_{\mu}^{T}(P)=\gamma_{\mu}-\frac{ \gamma \cdot P \;
 }{P^{2}} \, P_{\mu}$, $\gamma_\mu^A(P)=\sigma_{\mu\nu}P_{\nu}$ and $\bar{M}=\frac{2M_{f}M_{g}}{M_{f}+M_{g}}$. 
 
 Performing a Fierz transformation, it can be shown that the $\mathcal{K}^A$ piece, can be rewritten as $\frac{1}{3}\sigma_{\alpha\beta}\text{tr}_{D}[\sigma_{\alpha\beta}\chi_{H}(P)]$ and does not contribute to the PS and axial-vector channels. Conversely, the additional flexibility introduced by the parameter $\xi_A$ allows for a more accurate depiction of the vector meson mass spectrum; and, contrary to the RL alone (\emph{i.e.} $\xi_A=0$), it generates a  $F_V\neq 0$ component. As we will see later, this leads to improvements in the quark-photon vertex (QPV) and, consequently, in the description of electromagnetic processes\,\cite{Xing:2021dwe,Dang:2023ysl,Xing:2024bpj}. In line with a previous exploration of the $\gamma^* \gamma \to \pi^0$ process~\cite{Dang:2023ysl}, we have adopted the value $\xi_A=0.151$.

Note that, regardless of the truncation, the meson BSE can be cast as an eigenvalue equation by introducing a function $\lambda(P^2)$ in the right-hand-side of Eq.\,\eqref{eq:BSEGen}. Physical solutions lie at discrete values $\lambda(P^2_i=-m_i^2)=1$, such that the smallest $m_i$ corresponds to the ground-state meson mass. The resulting eigenvectors are associated with the BSAs, and must be canonically normalized according to~\cite{Nakanishi:1969ph}:
\begin{equation}\begin{aligned}\left[\frac{\mathrm{d}\ln\lambda(P^2)}{\mathrm{d}P^2}\right]_{P^2=-m^2_H}^{-1}=\mathrm{tr}\int_q\left[\bar{\Gamma}_{H}(-P)\chi_{H}(q;P)\right]\end{aligned},\end{equation}
where the $\text{tr}$ indicates trace over color, flavor and spinor indices. Omitting  the $\eta-\eta'$ mesons for now, in Table\,\ref{tab:Static} we collect some of the static properties of light pseudoscalar and  vector mesons. This list includes the corresponding decay constants, which are obtained from the following expressions:
\begin{eqnarray}P_\mu f_{PS}&=&\mathrm{tr}\int_q\left[\gamma_5\gamma_\mu S(q)\Gamma_{PS} S(q-P)\right]\,,\label{eq:decaycPS}\\
f_Vm_V&=&\frac{1}{3}\mathrm{tr}\int_q\left[\gamma_\mu S(q)\Gamma_\mu^V S(q-P)\right]\,.
\end{eqnarray}
\begin{table}[!htbp]
    \centering
        \caption{\label{tab:Static} Computed masses, BSAs and decay constants, $f_H$, for light PS ($\pi,K$) and V mesons ($\rho,K^{\star},\phi$). Mass units in GeV.}
    \begin{tabular}{c|c c c c}
        \hline
         & $m_H$ (GeV) &  $E_{H}$ & $F_{H}$&$f_{H}$ (GeV)\\
         \hline
        ~$\pi$~ & $0.140$& $3.595$& $0.475$& $ 0.101$ \\
        \hline
        ~$K$~ & $0.499$& $3.811$& $0.589$ &$ 0.106$\\
        \hline
        $\rho$& $0.879$& $1.442$ & $0.150$ &$ 0.190$\\
        \hline
        $K^{\star}$& $0.971$& $1.528$ & $0.184$ &$ 0.181$\\
        \hline
        $\phi$& $1.058$& $1.629$ & $0.231$ &$ 0.176$\\
        \hline
    \end{tabular}
    \label{vector}
\end{table}

\subsection{Quark-photon vertex}\label{sec:QPV}
The QPV satisfies the following inhomogeneous BSE:
\begin{eqnarray}
\label{eq:QPVinMRL}
\Gamma_\mu(Q)=\gamma_\mu-\frac{4}{3m_{G}^{2}}\int_{q}\left[\gamma_{\mu}\chi_\mu(Q)\gamma_{\mu} - \xi_{A} \tilde{\Gamma}_{j}\chi_H(P)\tilde{\Gamma}_{j}\right]\,,\hspace{0.6cm} 
\end{eqnarray}
where $\Gamma_\mu$ denotes the fully dressed QPV\,\footnote{For a specific quark flavor, $\Gamma_\mu^f$ must be weighted by the corresponding electric charge.}, $\chi_\mu$ is defined in analogy to Eq.\,\eqref{eq:BSWFdef} and, clearly, $Q$ is the photon momentum. Except for the inhomogeneity term $\gamma_\mu$, this equation resembles that of the vector meson, which entails the following structure for the quark-photon vertex ($\gamma_\mu^L=\gamma_\mu-\gamma_\mu^T$):
\begin{eqnarray}
\Gamma_\mu(Q)=f_L(Q^2)\gamma_\mu^L+f_T(Q^2)\gamma_\mu^T+\frac{1}{M}f_A(Q^2)\gamma_\mu^A\;.\,\label{eq:qfvstructure}
\end{eqnarray}
Plugging it into Eq.\,\eqref{eq:QPVinMRL}, it is revealed that $f_L(Q^2)=1$, guaranteeing that the longitudinal WGTI of the vertex be satisfied\,\cite{Xing:2022jtt}. For its part, the transverse components are:
\begin{eqnarray}
\label{eq:QPVdress1}
    f_T(Q^2)&=&\frac{\mathcal{I}}{\mathcal{I}-\tilde{\xi}\left[(C_{0})^{2}\hat{\xi}M^{2}+2\bar{C}_{0}\mathcal{I}\right]}\,,\\
\label{eq:QPVdress2}
    f_A(Q^2)&=&-\frac{\hat{\xi}M^{2}C_{0}}{\mathcal{I}-\tilde{\xi}\left[(C_{0})^{2}\hat{\xi}M^{2}+2\bar{C}_{0}\mathcal{I}\right]}\,;
\end{eqnarray}
where the $Q^2$-dependent quantities $\mathcal{I}$, $C_{\alpha}$ and $\bar{C}_{\alpha}$, are defined as follows:
\begin{equation}\begin{aligned}
\mathcal{I}&:=1-\hat{\xi}(2M^2C_0+C_2)\,,\\
C_{\alpha}&:=\int_{0}^{1}du\,I_{\alpha}(\omega_u(M^{2},Q^{2}))\,,\\
\bar C_{\alpha}&:=\int_{0}^{1}du\,u(u-1)I_{\alpha}(\omega_u(M^{2},Q^{2}))\,,\end{aligned}\end{equation}
with $\hat\xi=\frac{32\xi_{N}}{9m_{G}^{2}}$, $\tilde{\xi}=\frac{8Q^{2}}{3m_{G}^{2}}$ and $\omega_u = M^{2}+u(1-u)Q^{2}$.

These dressing functions are far more complex than $f_\mu^L$, but certain characteristics become apparent. Firstly, being associated with the $\gamma^T_\mu$ and $\gamma^A_\mu$ structures, both $f_{T,A}$ exhibit a vector meson pole on the timelike axis, namely $Q^2=-m_V^2$. Secondly, the QPV acquires the following form at $Q^2=0$:
\begin{equation}
    \Gamma_\mu(0)=\gamma_\mu + \frac{1}{M}f_A(0) \gamma_\mu^A  \,,
\end{equation}
implying $f_T(0)=1$ and that a non-zero $f_A(0)$ brings corrections to the otherwise bare vertex in this limit. Note this is achieved when $\xi_A\neq 0$, in which case a monotonically decreasing $f_A(Q^2)$ is produced. Consequently, being attached to $\gamma_\mu^A$, its presence is perfectly attributable to the emergence of the quark AMM\,\cite{Xing:2021dwe}. Another vital aspect of the QPV lies within its asymptotic profile:
\begin{equation}
    f_T(Q^2\to\infty)\to 1\,,\,f_A(Q^2\to\infty)\to 0\,,
\end{equation}
that permits to recover the corresponding point particle limit $\Gamma_\mu(Q^2\to\infty)\to \gamma_\mu$. An illustration of the $Q^2$ evolution of $f_{T,A}(Q^2)$ might be found in\,\cite{Xing:2021dwe}. For brevity, we restrain ourselves to quote their magnitude at $Q^2=0$ and slope,
\begin{equation}
\label{eq:defSlope}
    R_{T,A}:=-\frac{\partial f_{T,A}(Q^2)}{\partial Q^2}|_{Q^2=0}\,.
\end{equation}
As observed from the values in the Table\,\ref{tab:AMMval}, the AMM of the $s$-quark is lower than that of its lighter $u/d$ companions. Likewise, the slope of the dressing functions is less steep in this instance. Both observations are intuitive, merely indicating that the strange quark is heavier.
\begin{table}[!htbp]
    \centering
        \caption{\label{tab:AMMval}QPV dressing functions and their slope, Eq.\,\eqref{eq:defSlope}, at the soft-limit $Q^2=0$. Mass units of $R_{T,A}$ in GeV.}
    \begin{tabular}{c c c c c}
        \hline
        & $f_{T}^{f}$ & $f_{A}^{f}$ &  $R_{T}^{f}$ &  $R_{A}^{f}$  \\
        \hline
        $u/d$ & $1.0$ & $0.043$ & $0.49$& $0.056$ \\
        $s$ & $1.0$ & $0.053$ & $0.30$ & $0.047$ \\
        \hline
    \end{tabular}
    \label{QPV}
\end{table}

\section{Flavour mixing and the $\eta,\eta^{\prime}$ mesons}\label{sec:eta}
Due to the fact that $SU(2)$ flavor symmetry is a good approximation in nature, the pion is plainly decoupled from the $\eta-\eta'$ system\,\cite{Bhagwat:2007ha}, which incorporates the $s$-quark. The latter are in fact combinations of the octet and singlet states, $\eta_8-\eta_0$, so that the corresponding BSA of the mixed system can be written as:
\begin{equation}
\begin{aligned}
\Gamma_{\eta,\eta^{\prime}}(P)&=\mathcal{T}_0\,\Gamma_0^{\eta,\eta'}(P)+\mathcal{T}_8\,\Gamma_8^{\eta,\eta'}(P)\,,
\end{aligned}
\label{eq:etaBSA}
\end{equation}
such that the structure of $\Gamma_{0,8}$ is that of Eq.\,\eqref{eq:BSApi}, and $\mathcal{T}_{a}$ are the generators of the $U(3)$ group, which are associated with the flavor structure and satisfy $\text{tr}[\mathcal{T}_a \mathcal{T}_{b}]=\delta_{ab}$. Equivalently, the BSA might be also expressed in a flavor basis, in whose case:
\begin{equation}
\label{eq:etaBSA2}
    \Gamma_{\eta,\eta'}(P) = \mathcal{F}_l \,\Gamma_l^{\eta,\eta'}(P)+\mathcal{F}_s\, \Gamma_s^{\eta,\eta'}(P)\,,
\end{equation}
with $\mathcal{F}_l=\text{diag}(1,1,0)$ and $\mathcal{F}_s=\text{diag}=(0,0,\sqrt{2})$ expressing the flavor structure of the system. In line with the above, the corresponding BSWF is defined as follows:
\begin{equation}
    \chi_{\eta,\eta^{\prime}}(q;P) = \mathcal{S}(q)\Gamma_{\eta,\eta^{\prime}}(P) \mathcal{S}(q-P)\,,
\end{equation}
where $\mathcal{S}=\text{diag}[S_l,S_l,S_s]$. The RL truncation is incapable of producing flavor mixing and, coupled with the fact that $\mathcal{K}_A$ only affects vector channels, then the two-body kernel $\mathcal{K}_L$ must be supplemented to adequately describe these mixed states. Such an extension was proposed in\,\cite{Bhagwat:2007ha} and implemented throughout Refs.\,\cite{Ding:2018xwy,Raya:2019dnh,Zamora:2023fgl}; this reads:
\begin{equation}
\begin{aligned}
[\mathcal{K}_{N}]_{l_1 l_2}^{l_1^{\prime}l_2^{\prime}}&=\xi_{N}\frac{4}{3m_{G}^{2}}~\Big\{[\cos\theta_{N} i\gamma_5 z+\sin\theta_{N}\gamma_5\gamma\cdot \tilde{P} z]_{l_1^{\prime} l_2^{\prime}}\\ &\otimes[\cos\theta_{N} i\gamma_5 z+\sin\theta_{N}\gamma_5\gamma\cdot \tilde{P} z]_{l_1 l_2}
\Big\}\,,
\end{aligned}
\label{eq:kNA}
\end{equation}
where $\tilde{P}=P/M_l$, $z=\text{diag}[1,1,\nu_a]$ ($\nu_a$ being a parameter that introduces a dependence on $U(3)$ flavor-symmetry breaking); $\xi_N$ and $\theta_N$ control the strength of the NAA kernel and relative balance among its components, respectively. Here we have kept the indices $l_{1,2}^{(')}$, which describe the color and flavor structure, and capture the essence of the flavor mixing\,\footnote{For instance, the tensorial part of the RL kernel in Eq.\,\eqref{eq:K2CIRL}, would read $[\gamma_\mu]_{l_1 l_1'}\otimes [\gamma_\nu]_{l_2 l_2'}$, so it does not produce a mixture of flavors.}.

Thus, and remembering that the $\mathcal{K}_{\text{A}}$ kernel does not contribute to the pseudoscalar channels, the $\eta-\eta'$ mesons would be described by the following BSE:
\begin{equation}
[\Gamma_{\eta,\eta^{\prime}}(P)]_{l_1l_2}=\int_{q}[\mathcal{K}_{\text{L}}+\mathcal{K}_N]_{l_1 l_2}^{l_1^{\prime}l_2^{\prime}}(P)[\chi_{\eta,\eta^{\prime
}}(q;P)]_{l_1^{\prime}l_2^{\prime}}. 
\label{etaBSE}
\end{equation}
Substituting Eq.~(\Ref{eq:etaBSA}) into Eq.~(\Ref{etaBSE}), and properly projecting according to the spinor and flavor structure, one arrives at an eigenvalue equation of the form:
\begin{equation}
\lambda(P^{2})\begin{pmatrix}E_0\\F_0\\E_8\\F_8\end{pmatrix}=[\mathcal{K}(P^2)]_{4\times4}\begin{pmatrix}E_0\\F_0\\E_8\\F_8\end{pmatrix}.\end{equation}
Contrary to the previous cases, two ground-state solutions, $\lambda(P^2=-m_0^2)=1$, are found; the one producing the smallest $m_0$ is associated with the $\eta$ meson and, naturally, the heaviest one with the $\eta'$. In the case of ideal mixing, $\xi_N=0$, arising solutions could be understood as the non-physical $\eta_{8,0}$ states; or, in the flavor basis, as a pion-like system and a pure $s\bar{s}$ pseudoscalar. Along with $\theta_N$ and $\nu_a$, we fix $\xi_N$ to produce precise values $m_{\eta,\eta'}$. In practice, $\theta_N$ assumes small, nearly insignificant values, so fine-tuning does not result in any notable improvement. As a result, the $\gamma_5 \otimes \gamma_5$-proportional component encompasses nearly all the effects of the NAA kernel, meaning $\mathcal{K}_N$ is effectively controlled solely by the strength parameter $\xi_N$ and $\nu_N$. The kernel parameters, as well as the resulting masses and BSAs, are listed in the Table\,\ref{tab:massBSAeta}. 

\begin{table}[!htbp]
\caption{Masses (in GeV) and canonically normalized BSAs of the $\eta$ and $\eta^{\prime}$ mesons. The inputs entering the NAA kernel, Eq.\,\eqref{eq:kNA}, are: $\xi_N=0.2$, $\nu_N=0.71\approx M_l/M_s$ and $\theta_N=4^{\circ}$. 
    }
    \begin{tabular}{c|c c c c c c c c}
        \hline
        & Mass&$E_0$& $F_0$& $E_8$& $F_8$\\
        \hline
        $\eta$ & $0.549$& $0.297$& $0.169$& $3.644$& $1.282$ \\
        \hline
        $\eta^{\prime}$ & $0.958$& $2.373$& $0.935$& $-1.369$& $-0.461$ \\
        \hline
    \end{tabular}
    \label{tab:massBSAeta}
    \end{table}

In analogy to Eq.\,\eqref{eq:decaycPS}, we can also define decay constants $f_{8,0}^{H}$, obtained by the following formula:
\begin{equation}P_{\mu}f^{\eta, \eta^{\prime}}_{0/8}=\int_q \mathrm{tr}\left[T^H_{0/8}\gamma_{5}\gamma_{\mu}\mathcal{S}(q_{+})\Gamma_{\eta, \eta^{\prime}}\mathcal{S}(q_-)\right].\end{equation} These can be conveniently expressed in terms of two mixing angles $\theta_{8,0}$ as follows:
\begin{equation}\begin{pmatrix}f^\eta_0&f^\eta_8\\f^{\eta'}_0&f^{\eta'}_8\end{pmatrix}=\begin{pmatrix}-f_0\sin\theta_0&f_8\cos\theta_8\\f_0\cos\theta_0&f_8\sin\theta_8\end{pmatrix}.\end{equation}
The collection of masses and BSAs from  Table~\Ref{tab:massBSAeta} yield the following results: 
\begin{equation}
\begin{aligned}    f_0=0.162~\text{GeV}=1.60f^\pi&,~
    f_8=0.151~\text{GeV}=1.49f^\pi,
\\ \theta_0=-13.3^{\circ}&,~\theta_8=-16.5^{\circ},
    \end{aligned}
\end{equation} 
where the value $f_\pi=0.101$ GeV was obtained in a previous CI analysis of the $\gamma^* \gamma \to \pi^0$ transition,\,\cite{Dang:2023ysl}. For comparison, the DSE estimated from Ref.\,\cite{Ding:2018xwy} yields: $f_8=1.34f_\pi$, $f_0=1.24 f_\pi$ and $-\theta_{8,0}=0.21,\,2.8^{\circ}$; an estimate based on various phenomenological analyses is also reported therein, resulting in: $f_8=1.34(8)f_\pi$, $f_0=1.25(10) f_\pi$ and $-\theta_{8,0}=0.18(6),\,6(6)^{\circ}$.

A reliable estimate of the decay widths $\eta(\eta')\to\gamma \gamma$ is obtained from the corresponding decay constants, via the phenomenological formulae~\cite{Feldmann:1997vc}:
\begin{eqnarray}
\label{eq:deceta}
    \Gamma[\eta \to \gamma\gamma]&=&\frac{9\alpha^2}{16\pi^3} M_\eta^3 \left[\frac{C_8 f_{\eta^{\prime}}^0-C_0 f_{\eta^{\prime}}^8}{f_{\eta^{\prime}}^0 f_\eta^8-f_{\eta^{\prime}}^8 f_\eta^0}\right]^2 \,,\\
\label{eq:decetap}
    \Gamma[\eta^{\prime}\to\gamma\gamma]&=&\frac{9\alpha^2}{16\pi^3} M_{\eta^{\prime}}^3 \left[\frac{-C_8 f_\eta^0+C_0 f_\eta^8}{f_{\eta^{\prime}}^0 f_\eta^8-f_{\eta^{\prime}}^8 f_\eta^0}\right]^2\,;
\end{eqnarray}
where $C_0=(e_u^2+e_d^2+e_s^2)/\sqrt{3}$ and $C_8=(e_u^2+e_d^2-2e_s^2)/\sqrt{6}$. The arising estimations are captured in Table.\,\ref{tab:etaDecays}. These will be contrasted with the results produced in the impulse approximation in the following section.

\section{Radiative decays in the impulse approximation}\label{sec:raddecay}

\subsection{The $\eta(\eta') \to \gamma \gamma $ case}
The matrix element of the two-photon transition to a neutral pseudoscalar meson, $\mathcal{P} \to \gamma^*\gamma$, reads:
\begin{equation}
\label{eq:mathelemgg}
T_{\mu\nu}^{\mathcal{P}\gamma}(k_1;k_2)=g_{P\gamma}\frac{e^2}{(2\pi^2)^2}\epsilon_{\mu\nu\rho\sigma}k_1^{\rho}k_2^{\sigma}\,G_{\mathcal{P}\gamma}\left(k_1,k_2\right)\,;
\end{equation}
here $k_{1,2}$ denote the photon momenta, such that $(k_1+k_2)^2=-m_{\mathcal{P}}^2$; $e^2=4\pi \alpha_{em}$, with $\alpha_{em}\approx 1/137$ the electromagnetic coupling; and $g_{P\gamma}$ defines a coupling constant ensuring $G_{\mathcal{P}^0}(0,0)=1$. The decay $\mathcal{P} \to \gamma \gamma $ is defined in this soft limit and, consequently, the magnitude of $g_{\mathcal{P}\gamma}$ dictates the strength of the corresponding decay width,
\begin{equation}
\label{eq:decformg}
    \Gamma_{\mathcal{P}\gamma} = \frac{1}{4}\pi \alpha_{em}^2 m_\mathcal{P}^3\, |g_{\mathcal{P}\gamma}|^2\,.
\end{equation}

Capitalizing on the $\eta-\eta'$ case, the IA for the two-photon transition process yields\,\cite{Raya:2019dnh}:
\begin{equation}\label{eq:trianggg}
\begin{aligned}T_{\mu\nu}^{\mathcal{P}\gamma}(k_1;k_2)&=2 \,e^2\text{tr}\int_{q}\big[\mathcal{S}(q)i\mathbf{\Gamma}_{\mu}(-k_1)\mathcal{S}(q+k_1)\\&\times\Gamma_{\eta,\eta'}(k_1+k_2)\mathcal{S}(q-k_2)i\mathbf{\Gamma}_{\nu}(-k_2)\big]\,,
\end{aligned}
\end{equation}
where the trace is taken over color, spinor and flavor indices; the factor 2 appears in order to account for the possible ordering of the photons; and $\mathbf{\Gamma}_\mu=\text{diag}[2/3\, \Gamma_\mu^u,\, -1/3 \Gamma_\mu^d,\,-1/3 \Gamma_\mu^s]$ is defined in analogy to $\mathcal{S}$. Combining Eq.\,\eqref{eq:mathelemgg} and Eq.\,\eqref{eq:trianggg}, the computation of the form factor $G_{\mathcal{P}}$ follows after applying a proper projection operator. Nonetheless, certain issues concerning the definition of the $\gamma_5$ matrix emerge due to regularization-related aspects\,\cite{Korner:1991sx,Jegerlehner:2000dz,Belusca-Maito:2023wah}. This definition could affect the outcomes when dealing with an odd number of $\gamma_5$ matrices, as would be the case of the two-photon transition form factors. Following the analysis from Refs.\,\cite{Dang:2023ysl,Xing:2024bpj}, concerning the $\gamma^*\gamma \to \pi$ and $\gamma^* \to 3\pi$ anomalous processes in the CI model, we adopt the  representation suggested in\,\cite{Ma:2005md}:
\begin{equation}
    \gamma_5=-\frac{1}{24}\epsilon_{abcd}\gamma_{a}\gamma_{b}\gamma_{c}\gamma_{d}\,.
\end{equation}

As exposed in Eq.\,\eqref{eq:trianggg}, the IA permits the transition form factors to be expressed in terms of quark propagators, meson BSAs, and QPVs. The derivation of these pieces has been detailed throughout Sections\,\ref{sec:formalism} and \ref{sec:eta}. With the masses and BSAs collected in Table\,\ref{tab:massBSAeta}, the evaluation of Eqs.\,\eqref{eq:mathelemgg}-\eqref{eq:trianggg} yield the coupling constants and decay widths listed in Table\,\ref{tab:etaDecays}.
Evidently, the IA underestimates the experimental expectations on the $\eta(\eta')\to\gamma\gamma$ decay widths,\,\cite{ParticleDataGroup:2024cfk}. Such outcomes have been observed in other IA computations involving DSEs,\,\cite{Ding:2018xwy}, and extended Nambu-Jona-Lasinio frameworks\,~\cite{Takizawa:1996nw,Celenza:1999da}.  Furthermore, without considering the effects of the quark AMM, the decay widths in question would decrease by approximately $45\,\%$. This underscores the significance of this component in the QPV, which arises naturally within the present truncation. On the other hand, the results produced by the phenomenological formulae in Eqs.\,\eqref{eq:deceta}-\eqref{eq:decetap} yield values much closer to the empirically expected ones. Acknowledging all of these facts, the deficiencies of the present IA computation might be safely attributed to the failure of the triangle diagram in properly incorporating the beyond RL effects arising from the NAA. A symmetry-preserving derivation of such is a highly non-trivial task, \emph{e.g}\,\cite{Miramontes:2021xgn}.

\begin{table}[!htbp]
    \centering
          \caption{Coupling constants (in GeV$^{-1}$) and radiative decay widths (in keV) of the $\eta(\eta')\to\gamma\gamma$ processes;  impulse approximation computation (\emph{ia}) and Particle Data Group averages (\emph{pdg}), ~\cite{ParticleDataGroup:2024cfk}. The script $ph$ denotes the matching of the phenomenological formulae, Eqs.\,(\ref{eq:deceta},\ref{eq:decetap}), with Eq.\,\eqref{eq:decformg}.}
    \begin{tabular}{c|c c c c c c}
        \hline
        & $|g_{P\gamma}|^{ia}$ & $|g_{P\gamma}|^{ph}$ & $|g_{P\gamma}|^{pdg}$ & $\Gamma_{P\gamma}^{ia}$ & $\Gamma_{P\gamma}^{ph}$ & $\Gamma_{P\gamma}^{pdg}$  \\
        \hline
        $\eta\to \gamma \gamma$ & $0.242$ & $0.236$ & $0.273(5)$ & $0.405$ & $0.386$ & $0.515(18)$ \\
        \hline
        $\eta'\to \gamma \gamma$ & $0.273$ & $0.315$ & $0.343(6)$ & $2.744$ & $3.656$ & $4.34(14)$ \\
        \hline
    \end{tabular}
    \label{tab:etaDecays}
    \end{table}

\par

Anticipating the shortcomings of the CI in describing the $Q^2$ evolution of the form factors, and in particular the two-photon transitions\,\cite{Roberts:2010rn,Zamora:2023fgl}, results for $\gamma^{\star}\gamma\to \eta(\eta^{\prime})$ TFFs will not be shown. Instead, we limit ourselves to calculating the corresponding interaction radii, defined as follows:
\begin{equation}
\label{eq:radDef}
r_{M}^{2}:=-6\frac{1}{G(0)}\frac{d G(Q^{2})}{d Q^{2}}|_{Q^{2}=0}\,.
\end{equation}
Using the procedure outlined here, along with the corresponding model parameters, the resulting calculated values are:
\begin{equation}
r_{\eta}=0.50\,\text{fm}\, , r_{\eta^{\prime}}=\,0.46\,\text{fm}\, .
\end{equation}
Unsurprisingly, the individual interaction radii are underestimated (about $25\%$) in the CI model, as compared to the experimental data~\cite{A2:2013wad,BESIII:2015zpz,Escribano:2015yup,BESIII:2024pxo}. However, the computed ratio $r_{\eta}/r_{\eta^{\prime}}=1.09$ is plainly compatible with the empirical one: $r_{\eta}/r_{\eta^{\prime}}=1.08(7)$.\par

Having covered the treatment for the $\eta-\eta'$ mesons, along with its limitations,  we now move on to discussing the calculation of vector to pseudoscalar transitions.

\subsection{$\mathcal{V}\to \mathcal{P}\gamma$ decays}
The transition amplitude defining the $\gamma^*\mathcal{V}\rightarrow\mathcal{P}$ process adopts a similar form to that from Eq.\,\eqref{eq:mathelemgg}, namely:
\begin{equation}
\begin{aligned}T_{\mu\nu}^{\mathcal{V}\mathcal{P}}(k_1;k_2)&=e\,\frac{g_{\mathcal{VP}}}{m_{\mathcal{V}}}\,F_{\mathcal{V}\mathcal{P}}\left(k_1,k_2\right) \epsilon_{\mu\nu\rho\sigma}\,k_1^{\rho}k_2^{\sigma}\,,\end{aligned}\label{TFF}
\end{equation}
where $k_1$ and $k_2$ are the $PS$ meson and photon momenta, respectively, such that the on-shell conditions set: $k_1^2=-m_{\mathcal{P}}^2$, $k_2^2=Q^2$ and $(k_1+k_2)^2=-m_{\mathcal{V}}^2$. Thus, $F_{\mathcal{V}\mathcal{P}}$ depends solely on the photon momentum $Q^2$. The coupling constant $g_{\mathcal{V}\mathcal{P}}/m_{\mathcal{V}}$ is defined such that $F_{\mathcal{V}\mathcal{P}}(0)=1$, in whose case, the associated $\mathcal{V}\to \mathcal{P}\gamma$ decay width reads:
\begin{equation}
\Gamma_{\mathcal{V}\mathcal{P}}=\frac{\alpha_{\mathrm{em}}}{6m_{\mathcal{V}}}{\left(m_{\mathcal{V}}^2-m_{\mathcal{P}}^2\right)^3}\,\left[\frac{g_{\mathcal{V}\mathcal{P}}}{m_{\mathcal{V}}}\right]^2\,.
\end{equation}
Analogous to the two-photon decays, these processes can be evaluated in the impulse approximation, via:
\begin{equation}\begin{aligned}
T_{\mu\nu}^{VP\gamma}(k_1;k_2)&=e\,\mathrm{tr}\int_{q}\big[\mathcal{S}(q)\Gamma^{P}(-k_1)\mathcal{S}(q+k_1)\\
&\times\Gamma_{\mu}^{V}(k_1+k_2)\mathcal{S}(q-k_2)i\mathbf{\Gamma}_{\nu}(-k_2)\big]\,,\label{eq:PVtransition}
\end{aligned}\end{equation}
with the quark propagators, meson BSAs and QPVs determined as in previous sections.

The present framework leads us to the results shown in Table~\Ref{decay}, where both $\mathcal{V}\to\mathcal{P}\gamma$ decay widths and associated coupling constants are listed.  It is evident that our impulse approximation calculation yields radiative decay widths that are in reasonably good agreement with the empirical values\,\cite{ParticleDataGroup:2024cfk}, except for the $\phi\rightarrow\eta^{\prime}\gamma$ process.  Comparing the associated coupling constants shows much better agreement, suggesting that some of the discrepancy between the empirically estimated and calculated decay widths can be partially attributed to the computed vector meson masses. In this case, the level of compatibility in the 
$\rho^{\pm}\rightarrow\pi^{\pm}\gamma$ and
$K^{\star\pm}\rightarrow K^{\pm}\gamma$ cases is particularly notable. It is also observed the coupling constants $g_{\mathcal{\phi\eta}}$ and $g_{\mathcal{\phi\eta'}}$, as in the analogous two-photon transitions, turn out to be underestimated. Clearly, this is another measure of the incompleteness of the triangle diagram in incorporating corrections beyond RL and, without the consideration of the quark AMM in the interaction kernels, these constants would be suppressed by a further $25\, \%$. 

    \begin{table}[!htbp]
    \centering
          \caption{Dimensionless coupling constants and radiative decay widths (in keV) of vector-pseudoscalar radiative transitions; impulse approximation ($ia$) results and Particle Data Group averages ($pdg$),~\cite{ParticleDataGroup:2024cfk}.}
    \begin{tabular}{c|c c c c}
        \hline
        & $|g_{\mathcal{V}\mathcal{P}}|^{ia}$ & $|g_{\mathcal{V}\mathcal{P}}|^{pdg}$ & $\Gamma_{\mathcal{V}\mathcal{P}}^{ia}$ & $\Gamma_{\mathcal{V}\mathcal{P}}^{pdg}$ \\
        \hline
        $\rho^{\pm}\rightarrow\pi^{\pm}\gamma$& $0.32$& $0.36(2)$& $76.79$& $68(7$)\\
        \hline
        $K^{\star\pm}\rightarrow K^{\pm}\gamma$& $0.38$& $0.41(2)$& $64.50$& $50(5)$\\
        \hline
        $\phi\rightarrow\eta\gamma$& $0.30$& $0.34(1)$& $51.81$& $55(1)$\\
        \hline
        $\phi\rightarrow\eta^{\prime}\gamma$ & $0.34$ & $0.35(1)$ & $0.98$& $0.26(1)$\\
       
        \hline
    \end{tabular}
    \label{decay}
    \end{table}

In analogy with Eq.\,\eqref{eq:radDef}, we also calculate the corresponding interaction radii, obtaining:
\begin{eqnarray}\nonumber
    r_{\rho\to\pi}= 0.49\,\text{fm}&,&r_{K^*\to K}= 0.55\,\text{fm}\,,\\
    r_{\phi\to\eta}= 0.41\,\text{fm}&,&r_{\phi\to \eta'}= 0.42\,\text{fm}\,.
\end{eqnarray}
Referring to the value $r_\pi=0.455$ fm obtained within the CI framework,\cite{Sultan:2024hep}, we find $r_{\rho\to\pi}/r_\pi=1.08$. Meanwhile, the DSE calculation from Ref.,\cite{Maris:2002mz} produces $r_{\rho\to\pi}=0.69$ fm, leading to a ratio of $r_{\rho\to\pi}/r_\pi\approx 1.07$, which is in full agreement with the estimation made here.

\section{Summary}\label{sec:summary}
In this work, we have described a symmetry-preserving treatment of a vector$\times$vector CI model to examine the $\mathcal{V}\to\mathcal{P}\gamma$ and $\eta(\eta^\prime) \to \gamma\gamma$ radiative decays, addressing the shortcomings of the impulse approximation. The present approach extends previous literature by incorporating, simultaneously and rigorously, several elements. Among others, a regularization scheme capable of preserving all relevant symmetry constraints\,\cite{Xing:2022jtt}; and, the interaction kernels that take into account effects beyond the RL truncation\,\cite{Xing:2021dwe}. The latter allows the natural emergence of the anomalous magnetic moment of the quark (which does not happen in a RL treatment of CI), favoring the description of the vectorial channels and thus, the quark-photon vertex; in addition, the effects of the non-abelian anomaly are also considered, thus allowing a proper description of the mixed $\eta-\eta'$ states\,\cite{Ding:2018xwy,Zamora:2023fgl}.

In general, the static properties of the light pseudoscalar and vector mesons under examination are well discribed. This includes the mass spectrum and the corresponding decay constants. Furthermore, the impulse approximation computation of the  decay widths and coupling constants of the studied processes exhibit a reasonable agreement with the experimental determinations. Nonetheless, for the processes that involve the $\eta$ and $\eta'$ mesons, the results based on the impulse approximation tend to deviate more significantly from the empirical estimates of the corresponding decay widths. This is evident in the two-photon processes $\eta(\eta^\prime) \to \gamma\gamma$, by evaluating the corresponding decay widths using a phenomenological formula, which requires only knowledge of the masses and decay constants. In this scenario, the agreement between calculation and experiment is vastly better, suggesting a limitation of the IA: in the presence of the interaction kernel containing the non-Abelian anomaly, the triangle diagram arising from IA would be insufficient. This fact, briefly discussed in Ref.\,\cite{Ding:2018xwy}, has been overlooked in previous CI explorations.

Finally, it is worth emphasizing that despite the same approach has been effective in scrutinizing a collection of anomalous processes: $\gamma \gamma \to \pi^0$\,\cite{Dang:2023ysl}, $\gamma \to 3 \pi$\,\cite{Xing:2024bpj}, as well as the radiative decays $\mathcal{V}\to\mathcal{P}\gamma$ and $\eta(\eta^\prime) \to \gamma\gamma$ discussed in this work. In each instance, within the limitations of the CI, a good qualitative agreement with experimental observations and other theoretical approaches has been achieved, the significance of symmetries has been highlighted, and the good features and flaws of the framework have been plainly exposed. This highlights the valuable points of comparison the CI can provided, and establishes the effectiveness of the current approach for investigating these and many other type of quantities.

\begin{acknowledgments}
We acknowledges valuable discussions with Hao Dang and Zanbin Xing. Work supported by National Natural Science Foundation of China (grant no. 12135007). This work has also been partially funded by Ministerio Espa\~nol de Ciencia e Innovaci\'on under grant No. PID2019-107844GB-C22; Junta de Andaluc\'ia under contract Nos. Operativo FEDER Andaluc\'ia 2014-2020 UHU-1264517, P18-FR-5057 and also PAIDI FQM-370.
\end{acknowledgments}

\bibliography{ref}

\end{document}